\documentclass[sigconf]{acmart}

\usepackage{algpseudocode}
\usepackage[ruled,vlined,linesnumbered]{algorithm2e}
\usepackage{amsthm,amsmath}
\usepackage{marvosym}
\usepackage{graphicx}
\AtBeginDocument{
\providecommand\BibTeX{{
\normalfont B\kern-0.5em{\scshape i\kern-0.25em b}\kern-0.8em\TeX}}}

\setcopyright{acmcopyright}
\copyrightyear{2024}
\acmYear{2024}
\acmDOI{XXXXXXX.XXXXXXX}

\acmConference[KDD~'24]{Make sure to enter the correct conference title from your rights confirmation emai}{August 25--29, 2024}{Spain, ESP}

\acmPrice{15.00}
\acmISBN{978-1-4503-XXXX-X/18/06}

\begin{document}
\title{Category-Oriented Representation Learning for \\ Image to Multi-Modal Retrieval}

\author{Zida Cheng, Chen Ju, Shuai Xiao\textsuperscript{\Letter}, Xu Chen, Zhonghua Zhai, , \\ Xiaoyi Zeng, Weilin Huang, Junchi Yan}
\email{{chengzida.czd,huaisong.cx,zhaizhonghua.zzh,shuai.xsh,yuanhan,}@taobao.com,weilinh@hotmail.com,{ju\_chen,yanjunchi}@sjtu.edu.cn}
\affiliation{
\institution{Alibaba Group and Shanghai Jiao Tong University}
\country{China}
}

\renewcommand{\shortauthors}{Zida Cheng, et al.}

\begin{abstract}
The rise of multi-modal search requests from users has highlighted the importance of multi-modal retrieval (\textit{i.e.} image-to-text or text-to-image retrieval), yet the more complex task of image-to-multi-modal retrieval, crucial for many industry applications, remains under-explored. To address this gap and promote further research, we introduce and define the concept of Image-to-Multi-Modal Retrieval (IMMR), a process designed to retrieve rich multi-modal (\textit{i.e.} image and text) documents based on image queries. We focus on representation learning for IMMR and analyze three key challenges for it: 1) skewed data and noisy label in real-world industrial data, 2) the information-inequality between image and text modality of documents when learning representations, 3) effective and efficient training in large-scale industrial contexts.
To tackle the above challenges, we propose a novel framework named \textbf{o}rganizing \textbf{c}ategories and \textbf{le}arning by cl\textbf{a}ssification for \textbf{r}etrieval (\textbf{OCLEAR}). It consists of three components: 1) a novel category-oriented data governance scheme coupled with a large-scale classification-based learning paradigm, which handles the skewed and noisy data from a data perspective. 2) model architecture specially designed for multi-modal learning, where information-inequality between image and text modality of documents is considered for modality fusion. 3) a hybrid parallel training approach for tackling large-scale training in industrial scenario. The proposed framework achieves SOTA performance on public datasets and has been deployed in a real-world industrial e-commence system, leading to significant business growth. Code will be made publicly available.

\end{abstract}

\ccsdesc[500]{Information systems~Retrieval models and ranking}

\keywords{Image-to-multi-modal retrieval, Metric learning,  Modality fusion, Large-scale parallel training}

\maketitle

\section{Introduction}

In current Internet scenarios, there is a strong demand from users for searching multi-modal information, making multi-modal retrieval gain increasing attention. Among these studies, vanilla image-to-text or text-to-image retrieval have been well explored, but the more complex image-to-multi-modal-retrieval, prevalent in many industry applications, is largely ignored. For example, on an e-commerce platform (\textit{e.g.} Taobao), we can upload a photo of clothing, expecting the platform's algorithm to return the same or similar items containing both image and title information. Similarly, on social media, we may upload a photo of a tourist attraction, expecting to obtain photos of the same tourist attraction and textual comments from other users. Under these scenarios, we treat images as queries to accurately search for documents containing both images and textual descriptions. To promote community attention to this practical scenario, we formally define this task as \textbf{i}mage-to-\textbf{m}ulti-\textbf{m}odal-\textbf{r}etrieval~(\textbf{IMMR}), \textbf{which aims to find highly-similar multi-modal document given an image query}.


To advance the industrial IMMR task, this paper conducts pioneering explorations in this field. Representation for query images and multi-modal document is a key issue for IMMR task.
We review existing methods of related fields to analyze the challenges of representation learning for IMMR from three aspects, and correspondingly provide novel solutions:

\vspace{0.1cm}
1. \underline{Data Governance \& Learning Paradigm} 

In industrial retrieval scenarios, mainstream methods usually utilize pair-based learning paradigm: positive-negative pairs of samples are used to train the query-doc encoding networks, and binary classification loss~\cite{huang2013learning,lynch2016images} or metric learning loss\cite{kiros2014unifying,faghri2017vse++,wang2017adversarial,schroff2015facenet}~({\em e.g.}, triplet loss~\cite{schroff2015facenet}) are used as objective functions. 
The positive-negative pairs are usually constructed by user behaviour, {\em e.g.}, in e-commerce platform, a query and the clicked/unclicked item form a positive/negative pair. 
However, data collected in this way are usually \textbf{skewed and noisy}. 
For example, in e-commerce platform, different types of products are imbalanced in quantity. Clothing has significantly more queries and user behavior than furniture, making the collected data lacks sufficient samples for the furniture category. Thus, data skewness leads to poor representation learning of long-tailed products. More problematic, the randomness of user behavior also introduce substantial noise. For instance, users may click on only a small number of relevant items, thus many other relevant items are missed and incorrectly labeled as negative samples. 

Various methods in related fields has been developed for the issue of data skewness and noise. From the perspective of learning objective, some studies have focused on loss function to improve the model's learning ability. 
Both~\cite{focal} and~\cite{liu2022selfsupervised} explore loss re-weighting technique to improve learning balance on different instances. Recently, logit adjustment~\cite{hong2023long} has become a popular and promising technique for data imbalance problem. To tackle data noise, loss correction~\cite{Patrini_2017_CVPR,yaoyu,wang2021tackling} and loss reweighting ~\cite{wangruxin,chang2018active} techniques are designed. 
From the perspective of model architecture, \cite{jiang21a} designs an extra pruned branch to identify long-tail samples. DnC~\cite{DnC} trains individual experts on each subsets of dataset to enhance learning of long-tail samples.  To address data noise, a specific noise adaption layer is leveraged in \cite{chen2015webly,goldberger2017training} to estimate the noise transition pattern. \cite{han2018masking,tongxiao2015} propose to employ more dedicated model architecture(\textit{e.g.} GANs\cite{gans}) for robust learning.

Above methods have achieved remarkable performance in various public dataset and related scenarios. However, in our extremely large-scale industrial scenario, they still face several limits:

1) Existing methods generally explore the data skew and noise issues individually, but the two issues exist simultaneously in real-world scenario where existing methods can not handle both.
 
 2) Improvement from the perspective of model architecture more or less increases the inference complexity, which is not so efficient in large-scale industrial systems due to the strict latency requirement of online systems.

    3) In industrial scenarios, there exists natural additional information (\textit{e.g.}, inherent relationships between document images) that can help to build a more balanced and clean dataset from a data perspective, while previous approaches have largely ignored. Utilizing such additional information during the data organization is a more direct and efficient solution, compared to methods from perspective of learning objectives and model architecture.

In this paper, we propose a novel data governance and learning paradigm method, named \textbf{o}rganizing \textbf{c}ategories and \textbf{le}arning by cl\textbf{a}ssification for \textbf{r}etrieval (\textbf{OCLEAR}). It turns from pair-based to category-based learning. In particular, it contains a novel category-oriented data governance step and a large-scale classification learning step, and the latter is naturally derived from the former step. In data governance step, we organize query and document data into numerous fine-grained categories. Taking the e-commerce scenario as an example, we desire that one category consists of query images and items~(including item images and text) which contain the same product. We organize categories utilizing multiple sources of information, {\em i.e.}, natural relationships between item images on e-commerce platforms, user behavior and unsupervised clustering. In more detail, one item on e-commerce platform may contain multiple images thus they are naturally merged into one category; a query and the clicked item are merged; queries and items with high semantic similarity are merged by unsupervised clustering.

Such governance leads to a reduction in random noise, as we does not rely solely on user behavior. Data skewness is also significantly reduced, since cold docs which are rarely exposed to users can also be well utilized in our strategy, surpassing pair-based methods. 
With the large-scale fine-grained category data, we naturally consider the second step: large-scale fine-grained classification paradigm to learn query-doc representations for retrieval. For learning objective, instead of common softmax classification loss, we use cosine similarity and angular margin to enhance intra-class compactness and inter-class diversity and get better discriminative power.

\vspace{0.1cm}
2. \underline{Model Architecture for Modality Fusion in the Doc Side}

Existing cross-modal retrieval methods usually deal with only single-modal to single-modal retrieval, (\textit{e.g.} image-to-text/text-to-image retrieval)~\cite{lee2018stacked,chen2017amc,ju2022prompting,ju2023distilling,kiros2014unifying,yao2015learning,gan2016learning,ma2023diffusionseg,gordo2016deep}. IMMR task totally differs from them as the doc has both image and text modalities thus the modality fusion is important. Several excellent works have investigated on the multi-modality fusion problem\cite{li2020oscar, lu2019vilbert} for general domain data\cite{vilt,vlmo,blip,blip2}.
However, in industrial scenarios, the information-inequality between image and text modalities is a more important issue than general domain. That is, images provide extremely fine-grained details which is crucially important for retrieval, while textual descriptions encompass only high-level abstraction and concepts. Such characteristics of doc require us to design an appropriate modality fusion mechanism, to flexibly capture the modality interaction and better learn doc representations.

To tackle the unequal information of image-text modalities in the doc side, we design one concept-aware modal-fusion module, which consists of concept extraction and information fusion. In concept extraction, we employ external attention mechanism, because the shared K and V units within it are able to learn important high-level concepts from doc text. In information fusion, our insight is to make image play a major role and text act as a complementary role. Based on this insight, a cross attention mechanism from text to image is designed to fuse modalities, where text information serves as the Q of attention, while image information serves as K and V.  In the cross-attention procedure, the outcome of the attention is mainly affected by the V, while the Q plays the secondary role by influencing the weights multiplied with V.
Thus text acts as a complementary role by affecting weights. K and V are from image modality so that images still play the major role.

\vspace{0.1cm}
3. \underline{Efficient Training for Large-Scale Industrial scenarios}

In industrial contexts, there are stringent demands for computational efficiency and low latency, given that training datasets are often of an exceedingly large scale and inter-GPU communication overhead is subject to stringent upper-bound constraints.

To tackle the challenges of computational efficiency and minimal latency, we introduce a hybrid parallel training strategy, covering data parallelism and model parallelism. Data parallelism, as widely used in the industry, divides data into multiple batches and feed them to multiple GPUs simultaneously. 
Model parallelism is employed for the final classification layer to partition parameters across multiple GPUs, which reduces the GPU memory burden on each device for the large-scale fine-grained classification in our OCLEAR.
Furthermore, model parallelism could introduce the issue of inter-GPU communication as the gradients of the classification layer need to be exchanged among all GPUs. To reduce the communication overhead, we utilize K-nearest neighbor (KNN) softmax instead of the vanilla softmax, that is, allow for gradients to be exchanged only among limited GPUs.

\textbf{The contributions of this paper are summarized as follows:}

$\bullet$ We formally define the image-to-multi-modal-retrieval (IMMR) task where the query is an image, while the doc contains image and text description simultaneously, which is valuable for real-world industrial applications. 

$\bullet$ We propose one novel OCLEAR paradigm to tackle the IMMR task, consisting of a novel category-based data governance strategy, and large-scale fine-grained classification learning paradigm.

$\bullet$ We design a novel model architecture, where a concept-aware modality fusion module is proposed to adaptively fuse image-text information in the doc side. We propose a hybrid parallel training strategy to achieve efficient model optimization, with computational efficiency and low latency. 

$\bullet$ We conduct various experiments on both public benchmarks and large-scale industrial datasets. Results show that our framework performs SOTA on public benchmarks, and improves significantly in real-world search systems, which could inspire the subsequent community progress.

\section{Related Work}

\subsection{Cross-Modal Retrieval}
Most existing works for cross-modal retrieval focus on the scenarios where the query and the doc come from disjoint modalities~\cite{lee2018stacked,liu2023annotation,chen2017amc,ju2022prompting,ju2023distilling,kiros2014unifying,ju2023multi,yao2015learning,gan2016learning,ma2023diffusionseg,gordo2016deep,ma2023open,yang2023multi,liu2023audio}, such as image to text, or text to image. These methods can be grouped into two types: no-interaction and complex-interaction between queries and docs.

No-interaction methods can be further divided into two branches, that is, classification-based~\cite{huang2013learning,lynch2016images} and metric learning-based~\cite{kiros2014unifying,faghri2017vse++,wang2017adversarial}. The former uses two towers to represent queries and docs, and learns a pair-wise classifier by constructing positive-negative pairs. While the latter employs margin-based triplet loss~\cite{schroff2015facenet} for representation learning. For example, some large-scale pre-trainings use deep transformer, and follow the pair-based paradigm~\cite{CLIP,ju2023constraint,liu2022exploiting}. Kiros et al.\cite{kiros2014unifying} use deep convolutional neural networks to encode images and recurrent neural networks to encode sentences based on triplet loss. Wang et al.\cite{wang2017adversarial} use triplet loss plus auxiliary classification loss which differentiates image and text.

Complex-interaction methods~\cite{chen2017amc,lee2018stacked,diao2021similarity,li2019visual} leverage attention mechanism and graph matching for deep modeling between words and image regions. Oscar model~\cite{li2020oscar} is a transformer-based large-scale pre-training, where images and text are processed through one unified transformer architecture to understand complex interactions between query and doc. SGRAF~\cite{diao2021similarity} constructs graph by treating words and image regions as nodes, and learns interaction by graph propagation and graph attention.

\subsection{Deep Metric Learning}
Defining proper objectives can facilitate to distinguish objects' representations. In vision, deep metric learning has made a lot of progress. The contrastive paradigm, including contrastive loss~\cite{chopra2005learning}, triplet loss~\cite{schroff2015facenet}, quadruplet loss~\cite{chen2017beyond}, structured loss~\cite{oh2016deep}, etc, pulls a pair of samples together if their semantic labels are the same and push them away otherwise. Triplet loss takes a group of (anchor, positive, negative) as input and makes the anchor-positive distance smaller than the anchor-negative distance. 

Several issues exist for this paradigm: 1) A combinatorial explosion on the number of pairs or triplets with large-scale datasets, which leads to a significant increase in the number of training iterations to ensure the model has seen enough negative pairs. 2) Hard negative samples is important and informative for training models, however, hard sample mining is difficult and computationally expensive as the space of pairs is exponential~\cite{sohn2016improved}. Margin-based methods, such as Center loss~\cite{wen2016discriminative}, SphereFace~\cite{liu2017sphereface}, CosFace~\cite{wang2018cosface}, ArcFace~\cite{deng2019arcface}, etc, pull the features to corresponding class centers and require the angular distance between categories above a margin. SphereFace enforces the class centers far away from each other by introducing margins between categories. CosFace and ArcFace define the margin in the cosine space and angle space separately. Those methods have been demonstrated to work successfully in human face classification. The samples are only pulled to class centers so that the challenge due to the combinatorial explosion of pair samples in contrastive approaches are mitigated.

Inspired by this research line, we propose one class-centered learning paradigm for large-scale online cross-modal retrieval.

\section{Method} 
\noindent \textbf{Preliminaries}. \quad
Given an image $\mathbf{I}^{\mathrm{que}}\in \mathbb{R}^{H\times W\times 3}$ as the query, 
\textbf{i}mage-to-\textbf{m}ulti-\textbf{m}odal-\textbf{r}etrieval~(IMMR) aims to search for documents with similar semantics $\mathbf{D}$, that contain both images and textual descriptions. 
The IMMR model $\Phi_{\text{IMMR}}$ is formulated as:
\begin{align}
     \mathbf{D} = \Phi_{\text{IMMR}}(\mathbf{I}^{\mathrm{que}}) = \{\mathbf{I}^{\mathrm{doc}}\in \mathbb{R}^{H\times W\times 3}, \mathbf{T}^{\mathrm{doc}} \in \mathbb{R}^{T}\}_{i=1}^N, 
\end{align}
where $\mathbf{I}^{\mathrm{doc}}$ refers to doc-image, $\mathbf{T}^{\mathrm{doc}}$ refers to doc-text with $T$ words, and $N$ is the total number of retrieved doc entities. 
To tackle IMMR, an effective solution is to learn shared multi-modal space, and then retrieval by the embedding distance between queries and documents. 
And the most crucial aspect is the high-quality alignment of query-doc embeddings. We next propose one novel framework with effective data governance, model architecture \& training paradigm.

\begin{figure}[t]
\centering
\includegraphics[width=0.48\textwidth]{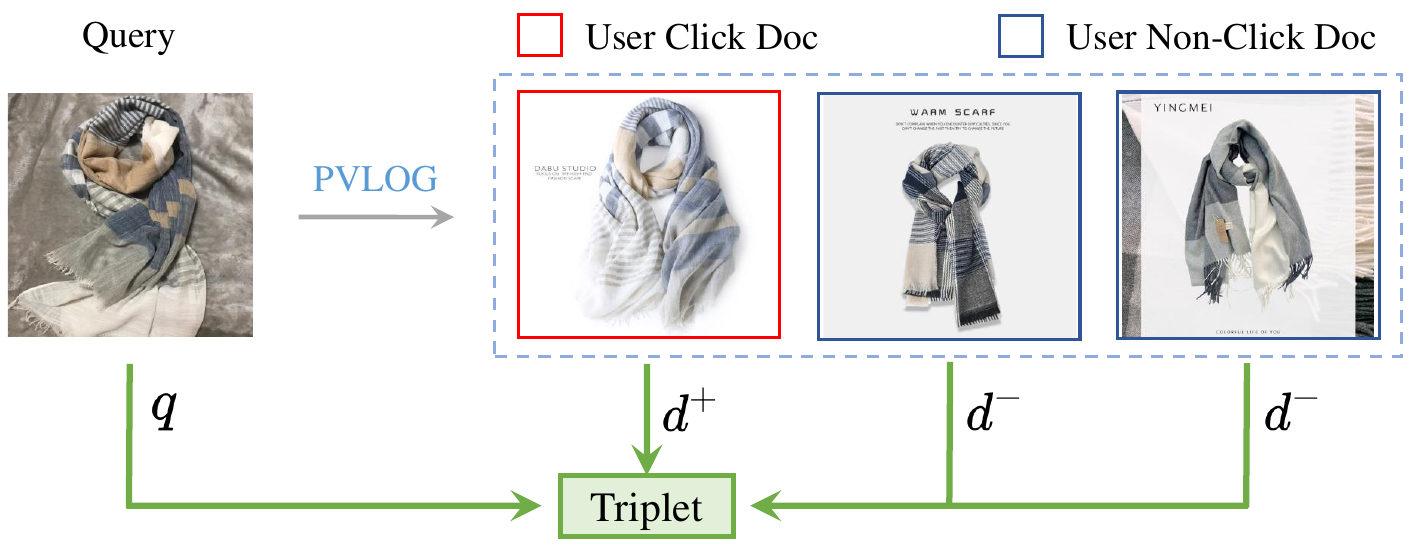}
\vspace{-15pt}
\caption{The industrial system exposure log and traditional training data collection paradigm. Red box is positive sample, blue box is negative sample.}
\vspace{-10pt}
\label{fig:triplet_data_collect}
\end{figure}

\subsection{Data Governance \& Learning}
To get well-aligned query-doc embeddings, one trivial idea is to follow existing industrial retrieval systems, {\em i.e.}, learning using \textit{pair-based data} from user behavior. As shown in Figure~\ref{fig:triplet_data_collect}, to construct training data, they refer to the system exposure log by composing (query, clicked item) as positive pairs, while (query, unclicked items) as negative pairs. Then, vanilla metric learning or binary classification are used to optimize models. 
However, \textit{pair-based data} usually suffer from two issues. \textbf{Skewness}: some items get larger probability of being exposed to users thus appear more frequently in data pairs. For example, in e-commerce scenarios, clothing types have significantly higher user queries than furniture or electronics, resulting in better/worse learning of popular/cold product types. \textbf{Noises}: as users' behavior has randomness, clicked items may be irrelevant to the query, while un-clicked ones may be relevant, leading to false labels. 
Under these two issues, pair-based learning faces great challenges to learn high-quality query-doc representations.

\begin{figure}[t!]
\centering
\includegraphics[width=0.45\textwidth]{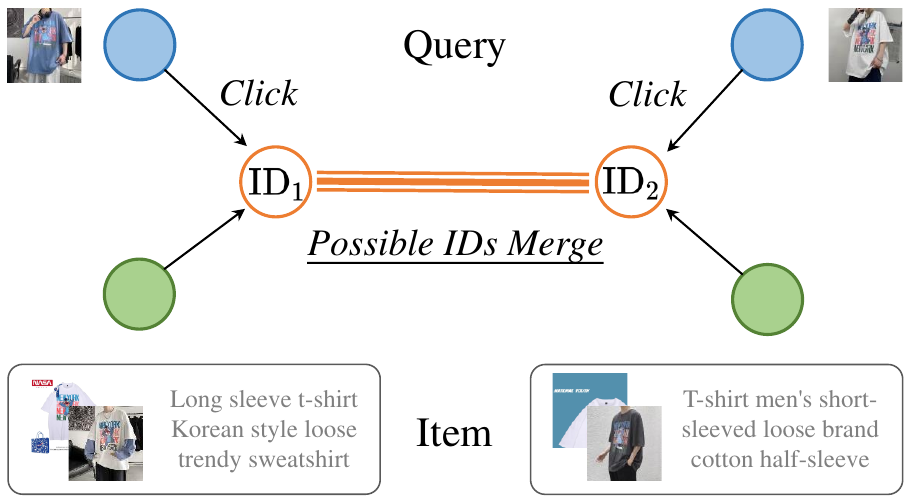}
\vspace{-10pt}
\caption{\textit{OCLEAR}: a novel data governance strategy, uses inherent doc information, user behaviors, and unsupervised clustering to merge data of the same category into one ID.}\vspace{-10pt}
\label{fig:new_data_collect}
\vspace{-0.2cm}
\end{figure}

\vspace{0.1cm}
To get rid of the above issues from existing industrial systems, we propose one novel idea for IMMR task named \textbf{o}rganizing \textbf{c}ategories and \textbf{le}arning by cl\textbf{a}ssification for \textbf{r}etrieval (\textbf{OCLEAR}), which incorporates one \textit{data governance step} to automatically organize multi-modal data into large number of categories, followed by embedding learning via \textit{large-scale fine-grained classification step}.

Below we first present the category-based data governance step, adopting IMMR in e-commerce scenarios as an example. Formally, we define the concept of category and samples:

\underline{\textit{Category}}: one category refers to one unique product, assigned with a unique ID. Note that, the same product may be sold by different retailers and may appear in different query images. Therefore, different items/query-images containing the same product should be classified into the same category. In following context, we use the terms "category" and "ID" interchangeably.

\underline{\textit{Sample}}: Each query image is considered as a sample, and each item image along with its corresponding textual description (a <item image, description text> tuple) is considered as a sample. In our actual system, we employ the item title as the description text. A category can contain multiple samples.

\vspace{0.1cm}
Figure~\ref{fig:new_data_collect} illustrates the organizational details of training data. Samples are grouped into categories using multiple sources of information, the procedure consists of three steps:

1. \textit{Merge doc samples by inherent information}. 
\hspace{1pt} In e-commerce platform, each item usually has multiple images (denoted as $\mathbf{I}_1,\mathbf{I}_2,\mathbf{I}_3,...$) and a title text ($\mathbf{T}$). Therefore, samples $<\mathbf{I}_1,\mathbf{T}>,<\mathbf{I}_2,\mathbf{T}>,<\mathbf{I}_3,\mathbf{T}>$ are assigned with the same category ID. 

2. \textit{Merge query and doc by user behavior}. 
\hspace{1pt} When one user initiates a query and clicks an item, the query is merged into the item's ID. 

3. \textit{Unsupervised clustering}. To perform unsupervised clustering, we firstly trained a ResNet-50 model with pair-based data by triplet-loss. Then we use it to compute the representation of query and item images within existing categories. Then for each existing category, we obtain the category prototype vector by averaging query and item image representations within it. Subsequently, using category prototype vectors, we perform unsupervised clustering. By adjusting the hyperparameters in the clustering process, categories containing the same products are grouped into the same cluster, thus these categories are merged into one new category.
\vspace{0.1cm}

After data governance, query-document containing the same products are organized into the same category. More exciting, the categories are fine-grained, the number of categories in real industrial datasets can reach tens of millions. Using such large-scale and fine-grained training data, we can optimize discriminative networks to learn well-aligned query-document embeddings, that distinguish subtle differences in multi-modal data.

With the above data, we naturally train one large-scale DNN model~(architectures detailed in next section) through the fine-grained classification paradigm. The goal is to enhance intra-class compactness and inter-class diversity, so that improving the model's discriminate power. 
Inspired by face recognition~\cite{deng2019arcface,wang2018cosface}, we here introduce angular margin enforcement in the loss function, instead of vanilla softmax cross-entropy. Principally, we utilize \textit{ID center proxy vector} (trainable) to represent the center of each category. Then in the loss function, we pull a sample's embedding close to its category proxy vector, and push it far away from other categories' proxies, where the distance is measured by cosine similarity. 
Formulaically, we denote the trainable center proxy vector of category $c$ as $\mathbf{w}_c$. For $i$-th sample~(query or doc), $y_i$ is its category label and $\mathbf{z}_i$ is its embeddings output from the DNN model. We denote the angle between category $c$'s proxy vector and $i$-th sample's embeddings as $\theta_{c,i} = <\mathbf{w}_c, \mathbf{z}_i>$. The fine-grained classification loss is:
\begin{equation}\label{equ:main_loss}
    \mathcal{L} = -\frac{1}{N}\sum_{i=1}^{N}\log \frac{\exp \{\mathcal{S}(\theta_{y_i, i}+m) / \tau\}}{\exp \{\mathcal{S}(\theta_{y_i, i}+m) / \tau\} + \sum\limits_{c\neq y_i} \exp \{\mathcal{S}(\theta_{c, i}) / \tau\}},
\end{equation}
where $m$ refers to the angular margin, $\mathcal{S}$ refers to cosine similarity, and $\tau$ is a temperature hyper-parameter. $m$ further enforces the embedding vectors of samples far away from other category-center proxies at least with a margin, enhancing intra-class compactness and inter-class diversity simultaneously.

\begin{figure*}[tb!]
\centering
\includegraphics[width=0.9\textwidth]{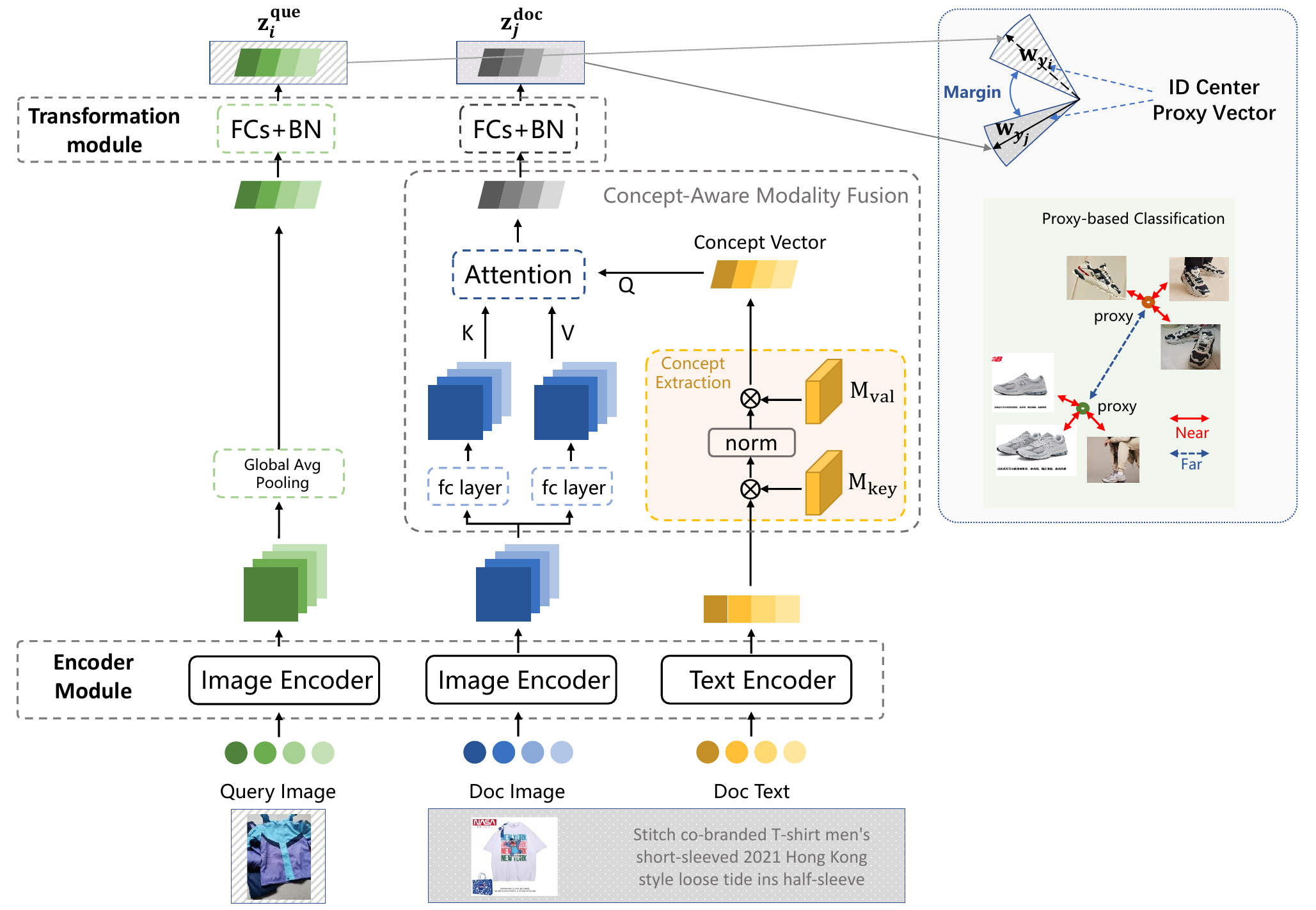}
\vspace{-10pt}
\caption{\textit{Model architecture consisting of four parts: image/text encoders, modality fusion, transformation module, and ID center proxies. In doc side, the concept-aware modal-fusion contains one concept extraction and one fusion network.}}
\label{fig:net}
\end{figure*}

\subsection{Model Architecture} 
As uni-modal query and multi-modal doc can have domain gaps ({\em e.g.}, doc images usually have higher resolution and purer background), we here design our model in the form of separate dual-towers. 
Figure~\ref{fig:net} shows the concrete architecture, covering four components: image/text encoders, concept-aware modality fusion module, transformation module, and ID center proxies.

For image/text encoders, query-images and doc-images are embedded by $\Phi_{\text{VE}}$, {\em e.g.}, ResNet~\cite{he2016deep} or Swin Transformer~\cite{liu2021swin}; while doc-texts are embedded by $\Phi_{\text{TE}}$, {\em e.g.}, BERT~\cite{devlin2018bert}. Formally, 
\begin{equation}
    {\mathbf{F}_{\mathrm{txt}}^{\mathrm{doc}} = \Phi_{\text{TE}}(\mathbf{T}^{\mathrm{doc}}), 
    \quad
    \mathbf{F}_{\mathrm{img}}^{\mathrm{doc}} = \Phi_{\text{VE}}(\mathbf{I}^{\mathrm{doc}}), 
    \quad
    \mathbf{F}_{\mathrm{img}}^{\mathrm{que}} = \Phi_{\text{VE}}(\mathbf{I}^{\mathrm{que}}), 
    }
\end{equation}
where $\mathbf{F}_{\mathrm{txt}}^{\mathrm{doc}} \in \mathbb{R}^{D}$, $\mathbf{F}_{\mathrm{img}}^{\mathrm{doc}} \in \mathbb{R}^{H\times W\times D}$, $\mathbf{F}_{\mathrm{img}}^{\mathrm{que}} \in \mathbb{R}^{H\times W\times D}$, and $D$ refers to the feature dimension.

\begin{algorithm}[tb!]
  \caption{Details of Concept Extraction $\Phi_{\text{CE}}$}
  \label{alg:concept extraction}
  \small
  \SetAlgoLined
  \KwIn{$D$: dimension of doc-text feature and output concept vector;\\
    \quad \quad \quad $E$: the number of columns in external \textit{key} unit; \\
    \quad \quad \quad $\mathbf{F}_{\mathrm{txt}}^{\mathrm{doc}} \in \mathbb{R}^{D}$: input doc-text feature from $\Phi_{\text{TE}}$; \\
    \quad \quad \quad $\mathbf{M}_{\mathrm{key}} \in \mathbb{R}^{E \times D}$: external \textit{key} unit, a trainable parameter; \\ 
    \quad \quad \quad $\mathbf{M}_{\mathrm{val}} \in \mathbb{R}^{D \times E}$: external \textit{value} unit, a trainable parameter; \\ 
  }
    Multiply input doc-text feature $\mathbf{F}_{\mathrm{txt}}^{\mathrm{doc}}$ with the external \textit{key} unit, then get one normalized weight vector by softmax: $\mathbf{w}_{\mathrm{cep}} = {\rm softmax}(\mathbf{M}_{\mathrm{key}} \mathbf{F}_{\mathrm{txt}}^{\mathrm{doc}}), \mathbf{w}_{\mathrm{cep}} \in \mathbb{R}^{E}$ \\
    
    Apply the weight vector $\mathbf{w}_{\mathrm{cep}}$ on the external \textit{value} unit, and get final concept vector: $\mathbf{c} = \mathbf{M}_{\mathrm{val}} \mathbf{w}_{\mathrm{cep}}, \mathbf{c} \in \mathbb{R}^D $. \\
  \KwOut{concept vector $\mathbf{c}$}
\end{algorithm}

\textbf{Concept-Aware Modality Fusion} $\Phi_{\text{CAMF}}$ is designed to fuse embeddings of doc-images and doc-texts. It includes two modules: concept extraction $\Phi_{\text{CE}}$ and fusion network $\Phi_{\text{FN}}$. Concept Extraction $\Phi_{\text{CE}}$ aims to get high-level semantics from the \textit{doc}-text modality. 
For input text features $\mathbf{F}_{\mathrm{txt}}^{\mathrm{doc}}$, it outputs a \textit{concept vector} $\mathbf{c} \in \mathbb{R}^{D}$. Please see \textbf{Algorithm 1} for more details. Structurally speaking, we adopt the \textit{external attention}~\cite{external-attention}, with two external memory units ($\mathbf{M}_{\mathrm{key}}$ and $\mathbf{M}_{\mathrm{val}}$) as key and value. $\mathbf{M}_{\mathrm{key}}$ and $\mathbf{M}_{\mathrm{val}}$ are trainable parameters and independent from specific input. Compared to vanilla attention, they are better at capturing the correlation between samples, as well as learning common concepts in text, {\em e.g.}, clothes' style and color on the e-commerce platform.

On the other hand, fusion network $\Phi_{\text{FN}}$ aggregates image-text modalities for final doc embeddings. As visual modality conveys detailed information while text conveys high-level abstract information, our design intuition is to let image plays a major role and text supplement it.
Structurally, we achieve the intuition via another attention procedure: K (key) and V (value) are acquired by applying two fully-connected layers on the doc-image feature, to ensure images still play the major role; while Q (query) is obtained with the \textit{concept vector} $\mathbf{c}$ from doc-texts, so that it only affects the weight multiplied with V, acing as a complementary role. In details, we first reshape the doc-image feature $\mathbf{F}_{\mathrm{img}}^{\mathrm{doc}} \in \mathbb{R}^{H\times W \times D}$ into dimension $HW \times D$; then apply fully-connected layers on the last dimension, to get K and V: $\mathbf{K}_\mathrm{img},\mathbf{V}_\mathrm{img} \in \mathbb{R}^{HW\times D}$; next we use concept vector $\mathbf{c}$ as Q to perform cross-attention calculation; finally the image-text fusion result $\mathbf{f}_\mathrm{fus} \in \mathbb{R}^D$ are achieved. 
\textbf{Algorithm 2} summarizes more details of fusion network.

\begin{algorithm}[tb!]
  \caption{Details of Fusion Network $\Phi_{\text{FN}}$}
  \label{alg:fusion}
  \small
  \SetAlgoLined
  \KwIn{$D$: dimension of fusion result vector ;\\
  \quad \quad \quad $\mathbf{F}_{\mathrm{img}}^{\mathrm{doc}} \in \mathbb{R}^{H\times W \times D}$: input doc-image feature from $\Phi_{\text{VE}}$;\\
  \quad \quad \quad $\textbf{c}$: concept vector from concept extraction $\Phi_{\text{CE}}$; \\

  \textbf{Trainable networks:} $\mathcal{F}_\mathrm{key}, \ \mathcal{F}_\mathrm{val}$ (two fully-connected layers)\\
  
  }
    Get Key and Value by applying the fully connected layers on the input doc-image feature: $\mathbf{K}_\mathrm{img} = \mathcal{F}_\mathrm{key} (\mathbf{F}_{\mathrm{img}}^{\mathrm{doc}}); \ \mathbf{V}_\mathrm{img} = \mathcal{F}_\mathrm{val} (\mathbf{F}_{\mathrm{img}}^{\mathrm{doc}})$; $\mathbf{K}_\mathrm{img},\mathbf{V}_\mathrm{img} \in \mathbb{R}^{HW\times D}$ \\

    Treat concept vector $\textbf{c}$ as Query, multiply it with $\mathbf{K}_\mathrm{img}$ and get the weight vector: $\mathbf{W}_\mathrm{fus} = {\rm softmax} (\mathbf{K}_\mathrm{img} \mathbf{c}),  \ \mathbf{W}_\mathrm{fus} \in \mathbb{R}^{HW}$ \\

    Apply weight vector on $\mathbf{V}_\mathrm{img}$ to get the fusion result: $\mathbf{f}_\mathrm{fus} = \mathbf{V}_\mathrm{img}^{\top} \mathbf{W}_\mathrm{fus}, \ \ \mathbf{f}_\mathrm{fus} \in \mathbb{R}^D$ \\
  \KwOut{image-text fusion result $\mathbf{f}_\mathrm{fus}$ on the doc side}
\end{algorithm}

After this, we consider to align uni-modal query and multi-modal doc into one shared embedding space. On the doc side, we convert $\mathbf{f}_\mathrm{fus}$ by transformation module $\Phi_{\text{DOC}}$, and obtained doc-embeddings $\mathbf{z}^\mathrm{doc}$; on the query side, we first perform global average pooling $\Phi_{\text{GAP}}$ on the query-image feature $\mathbf{F}_{\mathrm{img}}^{\mathrm{que}}$, then apply transformation module $\Phi_{\text{QUE}}$ to get query-embeddings as $\mathbf{z}^\mathrm{que}$. 
Formally,
\begin{equation}
    {\mathbf{z}^\mathrm{doc}= \Phi_{\text{DOC}}(\mathbf{f}_\mathrm{fus}) \in \mathbb{R}^{D}, 
    \quad
    \mathbf{z}^\mathrm{que} = \Phi_{\text{QUE}}\Phi_{\text{GAP}}(\mathbf{F}_{\mathrm{img}}^{\mathrm{que}}) \in \mathbb{R}^{D}. 
    }
\end{equation}

In the final, we introduce the classification layer with proxies: 
for each category (ID $c$), we set one trainable proxy vector $\mathbf{w}_c$, to act as the category center in the shared embedding space. Both query-embeddings $\mathbf{z}^\mathrm{que}$ and doc-embeddings $\mathbf{z}^\mathrm{doc}$ are forced to be close to their corresponding category proxy and far away from proxies of other categories. Note that, these proxy vectors are only used for training, and we can remove them during inference.

\begin{figure}[t!]
\centering
\includegraphics[width=0.49\textwidth]{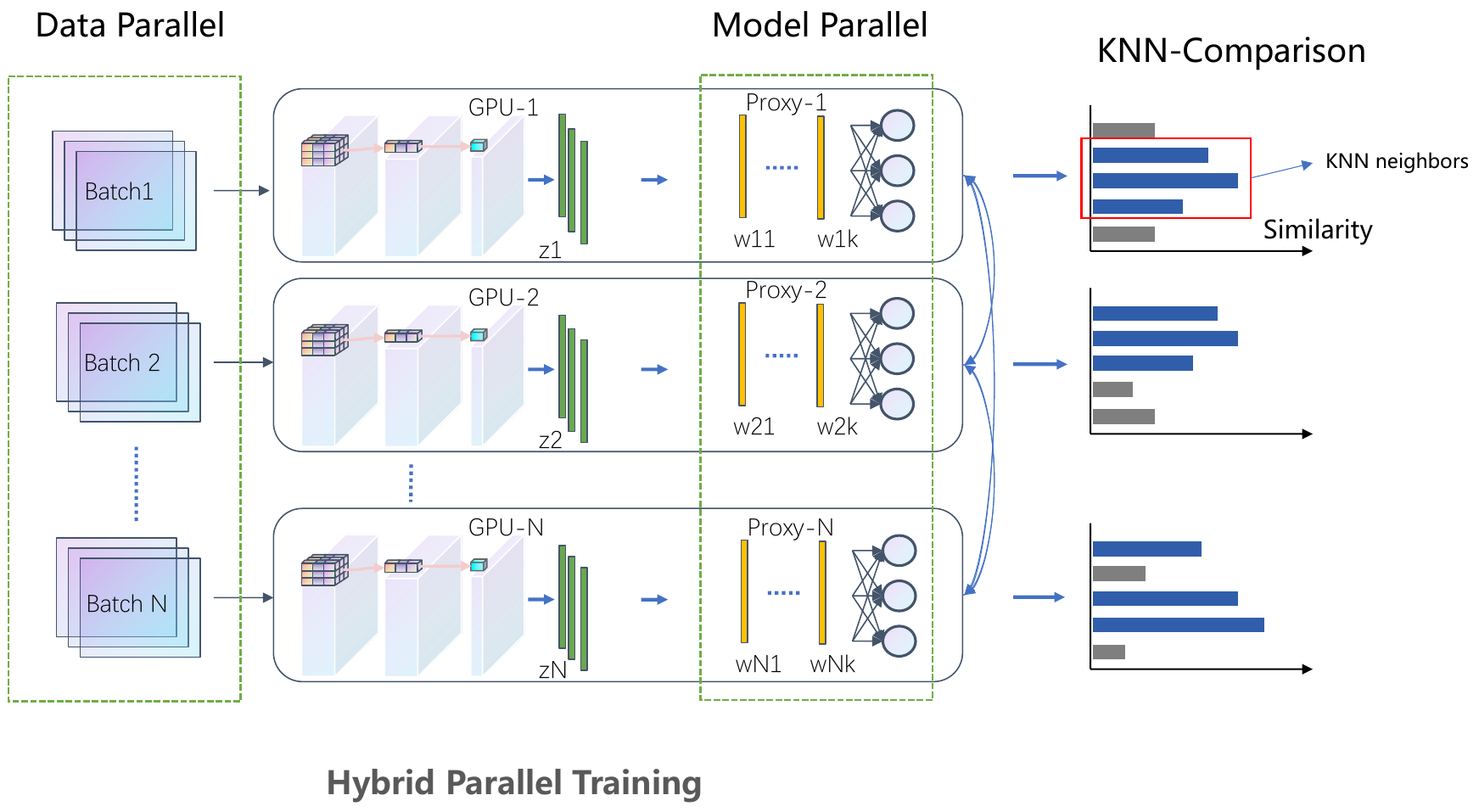}
\vspace{-10pt}
\caption{Hybrid parallel training consists of data parallel, model parallel, and an efficient implementation of final classification layer by KNN-comparison.}\vspace{-10pt}
\label{fig:hybird training}
\end{figure}

\subsection{Hybrid Parallel Training} 
In the industrial scenarios, training our model faces three challenges: large-scale data, significant number of model parameters, and inter-GPU communication overhead.  Shown in Figure~\ref{fig:hybird training}, we design a hybrid parallel training strategy for the above challenges, consists of data parallel, model parallel and an efficient implementation of final classification layer by KNN-comparison.

For data parallelism, we divide data into multiple batches and feed them to multiple GPUs simultaneously, as the widely-used solutions in the industrial setting. 

Model parallelism is utilized to handle the proxy vectors in the final classification layer. Due to large number of categories, the proxy vectors lead to large number of parameters. Thus we split the proxy parameters into multiple GPUs, given that the parameters of proxies can be larger than the memory capacity of one GPU.

KNN-comparison aims to tackle inter-GPU communication issue. To get final loss function, we need to compute the cosine similarities between samples' embedding vectors and proxies located at different GPUs. A large amount of inter-GPU communication take place here. To reduce the communication cost and speed up training, we use KNN-comparison method inspired by~\cite{song2020large}. For each sample of category $c$, only top-K similar proxies to proxy $\mathbf{w}_c$ are used for loss calculation in Equation~\ref{equ:main_loss}, as shown in Figure~\ref{fig:hybird training}. The $K$ is set to $10\%$ of the number of categories. We maintain a IDs proxies similarity matrix using cosine distance between proxies to calculate the top-$K$ similar proxies and the matrix is updated during training.

\section{Experiment}
We evaluate on public real-application datasets and compare with SOTA methods. Meaningful discoveries are discussed.

\subsection{Dataset}
\begin{figure}[t!]
\centering
\includegraphics[width=0.4\textwidth]{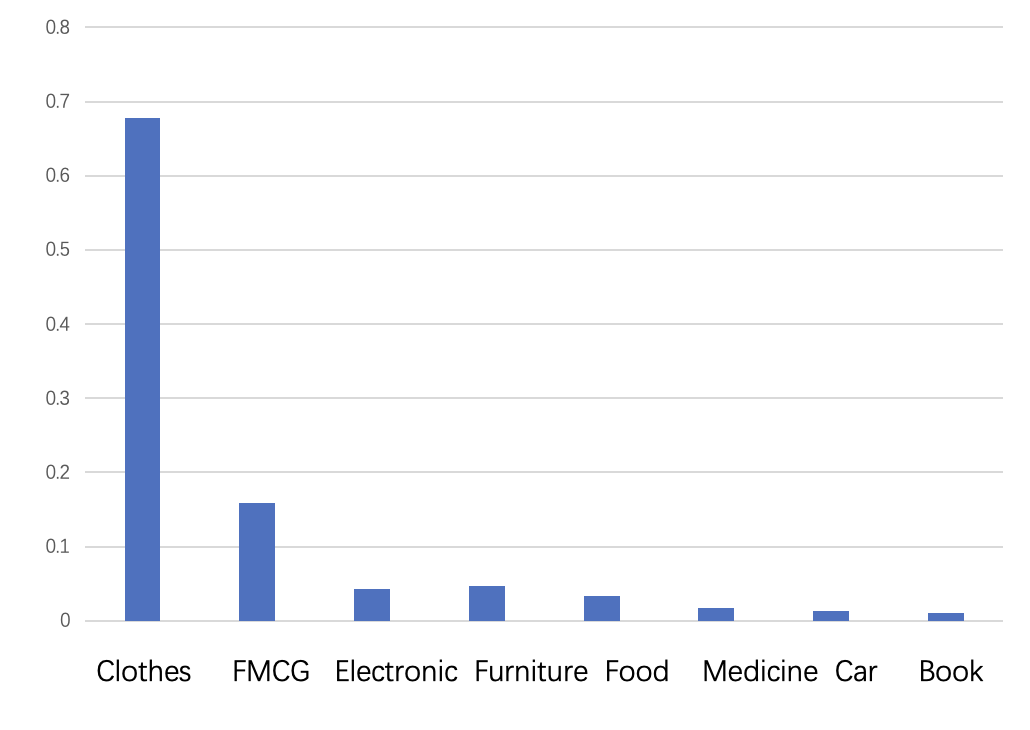}
\vspace{-10pt}
\caption{Statistical distribution of exposure categories.}\vspace{-10pt}
\label{fig:distribution}
\end{figure}

\textbf{AliProduct} is a large-scale and fine-grained product dataset collected from Alibaba e-commerce platform~\cite{cheng2020weakly}~\footnote{https://tianchi.aliyun.com/competition/entrance/531884/information}. The dataset covers different categories, such as clothes, cosmetics, foods, devices, toys and other fast-moving consumer goods (FMCG). As the distribution of exposure shown in Figure~\ref{fig:distribution}, this dataset is skewed. Each product has several images, one title, and also query images which can be linked to products by user click. Overall, the training set includes $12$ million products with $1.1$ billion images and $12$ million texts. The validation set has $240$ thousand positive-negative pairs of (queries, docs). The test set has $1500$ queries, each of which has $100$ candidate products with labels indicating their relationship to queries.

\noindent \textbf{Taobao Live} refers to a multi-modal dataset named Watch and Buy (WAB)~\footnote{https://tianchi.aliyun.com/dataset/dataDetail?dataId=75730} from Alibaba's Taobao livestreaming platform. During the broadcast, the live streamer displays lots of clothes, which are sold at Taobao. By annotating keyframes in videos and linking to products, it is beneficial for customers if the product being displayed by streamers is identified and recommended in time. For each product, the product in images is annotated and product title text is provided. Totally, $1,042,178$ images are annotated. The dataset is split into 70\% for training, 10\% for evaluation, and 20\% for testing.

\subsection{Experiment Details}
For images in the AliProduct dataset, we use YOLO-based method~\cite{ redmon2018yolov3} to detect the main object of interest, then crop the object region and finally resize the images to $224 \times 224$. For the Taobao Live dataset, we use the given bounding boxes to crop images and finally resize them to $224 \times 224$; we clip the product titles with a maximum length of 20, which is adequate for most sentences. For the doc~(product) side, we treat each object/box as an independent entity. We use ResNet50~\cite{he2016deep} as image encoders for both query and doc, while BERT~\cite{devlin2018bert} to get text embedding of doc. For compared approaches, we use the same feature extractors. We use SGD, with a learning rate of $0.001$. The ResNet image encoder is pre-trained on ImageNet~\cite{ImageNet} and other parts are trained from scratch.

For concept-aware modality fusion, we set the final dimension of concept vector as $D=256$. In the external memory mechanism, there are $E$ rows/columns in $\mathbf{M}_\mathrm{key}/\mathbf{M}_\mathrm{val}$, we set $E=16$. In the classification loss (Eq.~\ref{equ:main_loss}), we set the temperature hyper-parameter$\tau=64$, and the angular margin $m=0.5$.

\begin{table}[t!]
\caption{Comparison of image-to-multimodal retrieval on AliProduct dataset in terms of Precision and MAP/MRR.}
\vspace{-10pt}
\label{tab:aliproduct_results}
\begin{tabular}{crrrr}
\toprule
Approach & Identical@1 & Relevance@1 & MAP &MRR\\
\midrule
\textit{DSSM} & 66.58\% & 73.46\% & 0.5283 & 0.6973\\
\textit{Triplet} & 68.53\% & 71.36\% & 0.5312 & 0.7015\\
\textit{DSSM-C} & 66.92\% & 69.83\% & 0.5286 & 0.6989\\
\textit{Ensemble} & 72.47\% & 76.25\% & 0.5562 & 0.7683\\
\textit{CLIP} & 71.68\% & 73.38\% & 0.5462 & 0.7425\\
\textit{Oscar} & 69.03\% & 71.24\% & 0.5378 & 0.7153\\
\textit{ViLBERT} & 69.61\% & 71.89\% & 0.5403 & 0.7282\\
\textit{Ours-I} &75.82\% & 74.68\% & 0.5875 & 0.7983\\
\textit{Ours-E} &75.61\% & 77.28\% & 0.5812 & 0.7942\\
\textit{\textbf{Ours}} &\textbf{78.41}\% & \textbf{79.82}\% & \textbf{0.6075} & \textbf{0.8164}\\
\bottomrule
\end{tabular}
\vspace{-0.8cm}
\end{table}

\noindent \textbf{Compared Approaches:} 

(1) Deep Structured Semantic Model (\textit{DSSM})~\cite{huang2013learning,lynch2016images} computes the inner-product similarity between queries and docs, then conducts binary classification. We uses ResNet50 for query and doc image here, while adopts BERT for doc text, and fuses the doc embedding of image and text by average pooling.

(2) \textit{Triplet} loss~\cite{schroff2015facenet}. Here the anchor is the query image, and the positive/negative instance is the item image plus text that has the same/different label with queries. The embedding of doc image-text is combined by average pooling.

(3) \textit{DSSM-C}~\cite{wang2017adversarial} is DSSM plus auxiliary classification loss. We use coarse-grained category labels of Taobao Live~(only 23 categories), and append (a fully connected layer \& softmax) to encoders for cross-entropy classification. This setting aims to study classification ablation between coarse-grain and large-scale fine-grain.

(4) \textit{Ensemble} refers to concatenating the embedding from Triplet and DSSM, then evaluating the corresponding performance.

(5) \textit{CLIP}~\cite{radford2021learning} is the two-tower vision-language foundation model. By pre-training on large-scale image-text pairs from web, it shows powerful “zero-shot” capability. \textit{CLIP} is data-hungry and computationally cost, where the training data is $0.4$ billion image-text pairs and the optimization uses 256 GPUs for 2 weeks.

(6)~Oscar~\cite{li2020oscar} uses salient object tags detected in images as anchor points to align with text descriptions. The model is pretrained on the public corpus of 6.5 million text-image pairs.

(7)~ViLBERT~\cite{lu2019vilbert} follows a BERT-based model for learning task-agnostic joint representations of image content and language. ViLBERT firstly encodes both visual and textual inputs in separate towers and then interacts via co-attentional transformer layers.

(8)~BLIP~\cite{blip} leverages noisy web data by bootstrapping text captions. BLIP designs a captioner to generate synthetic captions and a filter to remove the noisy ones.

(9) \textit{Ours-I} and \textit{Ours-E} are degraded versions of our method for ablations. In \textit{Ours-I}, only the image is used on the doc side, to study the impact of the text. In \textit{Ours-E}, the final doc embedding is obtained by simply averaging the image and text embedding, as used to study the impact of the concept-aware modality fusion.

\noindent \textbf{Evaluation Metrics.} 
For inference, we calculate the embedding similarity between query and doc, and sort the doc by similarity scores. Note that, one query may correspond to multiple correct results. Identical@$k$ refers to the ratio of queries where the identical product appears in the top $k$ results. Relevance@$k$ is similar, but it considers human-labeled products which are relevant but not identical to the query. Mean average precision~(MAP) and mean reciprocal rank~(MRR) are also used. For MRR, we consider the first correct product in the retrieved result.

\begin{table}[t!]
\caption{Comparison of image-to-multimodal retrieval on Taobao Live dataset in terms of Precision and MAP/MRR.}
\vspace{-10pt}
\label{tab:alilive_results}
\begin{tabular}{crrrr}
\toprule
Approach & Identical@1 & Identical@5 & MAP &MRR\\
\midrule
\textit{DSSM} & 44.70\% & 51.48\% & 0.3338 & 0.4769 \\
\textit{Triplet} & 46.79\% & 55.36\% & 0.3686 &  0.4831\\
\textit{DSSM-C} & 41.09\% & 53.66\% & 0.3103 & 0.4597 \\
\textit{Ensemble} & 49.12\% & 59.49\% & 0.3959 & 0.5288 \\
\textit{CLIP} & 47.94\% & 57.37\% & 0.3846 & 0.5192 \\
\textit{Oscar} & 46.89\% & 55.93\% & 0.3701 & 0.4914 \\
\textit{ViLBERT} & 47.02\% & 56.13\% & 0.3725 & 0.5087 \\
\textit{BLIP}   & 50.18\% & 58.84\% & 0.3912 & 0.5282 \\
\textit{Ours-I} & 51.16\% & 66.33\% & 0.4022 & 0.5852 \\
\textit{Ours-E} & 51.01\% & 66.23\% & 0.4040 & 0.5872\\
\textit{\textbf{Ours}}  & \textbf{52.86}\% & \textbf{67.48}\% & \textbf{0.4148} & \textbf{0.5971}\\
\bottomrule
\end{tabular}
\end{table}

\begin{figure*}[t!]
\centering
\begin{minipage}[t]{0.48\textwidth}
\centering
\includegraphics[width=\textwidth]{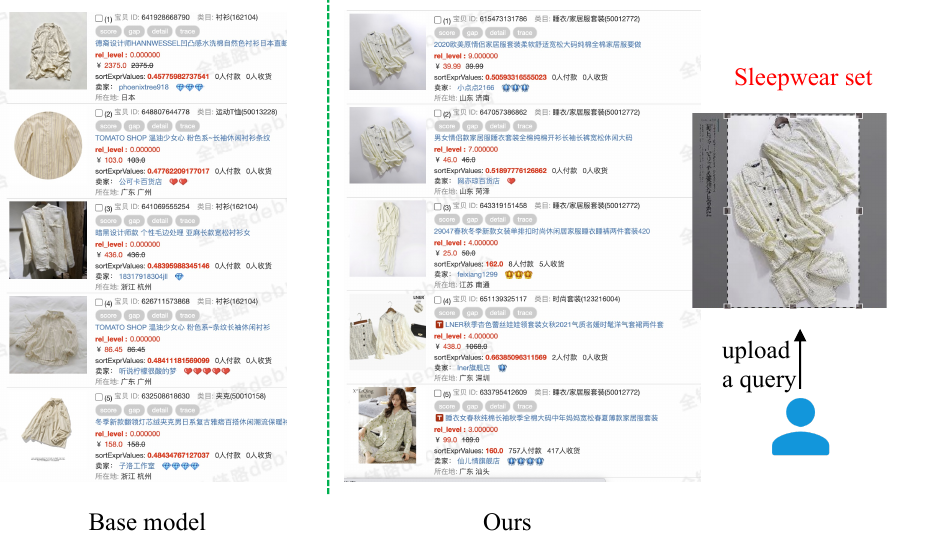}
\vspace{-0.6cm}
\caption*{(a) \footnotesize{example of sleepwear set}}
\end{minipage}
\hspace{5pt}
\begin{minipage}[t]{0.48\textwidth}
\centering
\includegraphics[width=\textwidth]{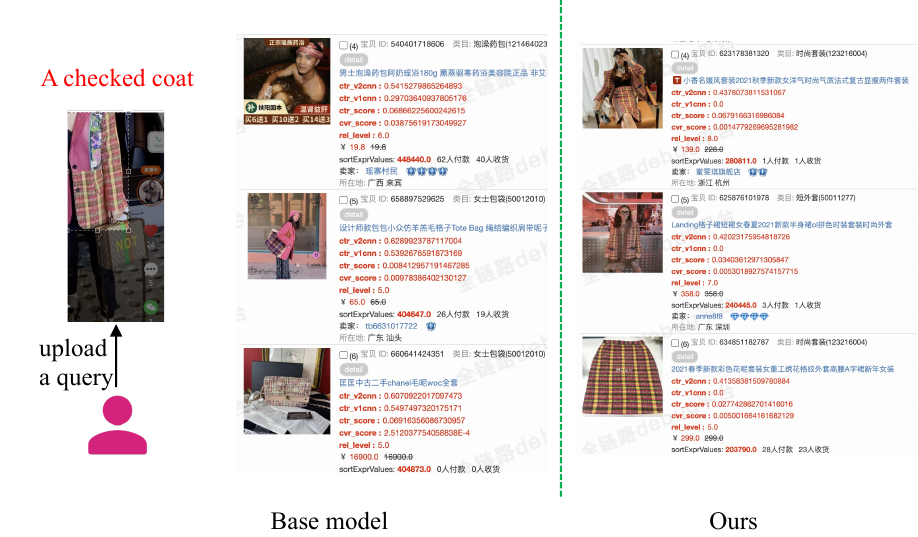}
\vspace{-0.6cm}
\caption*{(b) \footnotesize{example of a coat}}
\end{minipage}
\vspace{-10pt}
\caption{Ranking examples of online Base model \textit{v.s.} our model. Given an image query, we first use online detectors to localize the bounding box, then list the ranking items of different models. In (a), both models give relevant results, but ours ranks items containing the same product as the query in the top-2 position. In (b), the query is a hard case as only half of the coat appears, our model returns relevant products while the Base model gives irrelevant results such as bags.}
\label{figure:ranking_exmples}
\vspace{-0.2cm}
\end{figure*}

\subsection{Results \& Discussion}
\underline{Q1: How much performance can the proposed method improve?}

\vspace{0.05cm}
As shown in Table~\ref{tab:aliproduct_results} and Table~\ref{tab:alilive_results}, our method outperforms alternatives with a notable margin on both dataset. Since most of the products in the dataset are clothes, this is a fine-grained scenario and our method has advantage. In general, with the OCLEAR learning paradigm and concept-aware modal-fusion, our method outperforms several main-stream retrieval methods significantly.

\vspace{0.15cm}
\noindent \underline{Q2: What can text contribute to cross-modal retrieval?}

\vspace{0.05cm}
As shown in Table~\ref{tab:aliproduct_results},  Relevance@1 is improved a lot when text is considered~(\textit{Ours} v.s. \textit{Ours-I}, \textit{Ours-E} v.s. \textit{Ours-I}). Using concept-aware modal-fusion, Identical@1 also improves significantly~(\textit{Ours} v.s. \textit{Ours-I}). In Taobao Live dataset of Table~\ref{tab:alilive_results}, we still see a gain of Identical@x and MAP/MRR. In general, text information can improve the overall relevance of retrieval to the query.

\vspace{0.15cm}
\noindent \underline{Q3: What impact can concept-aware modal-fusion bring?}

\vspace{0.05cm}
As shown in Table~\ref{tab:aliproduct_results} and Table~\ref{tab:alilive_results}, \textit{Ours} outperformes \textit{Ours-E}, especially on the large AliProduct dataset. Since the image and text modalities are unequal for doc, naive modality fusion methods,e.g. average pooling in \textit{Ours-E}, can not perform well. The concept-aware modal-fusion first extracts concept from the text by external memories and fuse two modalities by attention mechanism, which is more adaptive and reasonable.

\vspace{0.15cm}
\noindent \underline{Q4: Classification comparison between fine-grain to coarse-grain?}

\vspace{0.05cm}
By comparing \textit{Ours/Ours-E} and \textit{DSSM-C}, we can see fine-grained classification loss is better for large-scale retrieval tasks. Because the fine-grained category label naturally meets our needs for discriminating subtle differences between samples. While the coarse-grained label provides little helpful information. Besides, see Table~\ref{tab:alilive_results}, as the products in Taobao Live is more fine-grained, we can see adding coarse-grained classification loss is even harmful by comparing \textit{DSSM-C} and \textit{DSSM}.

\begin{table}[tb!]
\caption{Ablation studies about the number of training samples in one category on AliProduct dataset~(Identical@1).}
\vspace{-10pt}
\label{tab:taobaolive_ablation}
\centering
\begin{tabular}{cclll}
\toprule
Sample Size & 5 & 10 & 15\\
\midrule
Identical@1  & 65.42\% & 74.85\% & 78.41\% \\
\bottomrule
\end{tabular}
\end{table}

The proposed large-scale classification paradigm for multi-modal retrieval relies on adequate samples in large-scale categories. We conduct experiments to study the number of samples in one category in  AliProduct, where the maximum number of samples per category is clipped. See Table~\ref{tab:taobaolive_ablation}, the retrieval performance improves significantly when more samples are accessible in one category.

\begin{table}[tb!]
\centering
\caption{Overall comparison results on real industrial scenario. Identical and Relevance are offline evaluation metric indicates the ratio of retrieved results relevant/identical to the query content. CTR, CVR and deal number are online A/B test metrics. Due to company's confidentiality regulations, online deal number of base model are blinded and denoted as $\star$, but we provide the improvement gap.}
\vspace{-8pt}
\label{table:overall_industrial}
\renewcommand{\arraystretch}{1.0}
\setlength{\tabcolsep}{0.8mm}{ 
\scalebox{0.9}{
\begin{tabular}{cccclc}
\hline
Model  & Identical &  Relevance   & CTR          &CVR& deal number    \\ \hline
Base   & 71.93\%   & 73.48\%      &   9.30\%    &2.01\%& $\star$      \\
Ours   & \textbf{82.14}\%   & \textbf{80.78}\%      & \textbf{9.84}\%    & \textbf{2.10}\%& \textbf{+6.45}\% \\ \hline
\end{tabular}
}}
\end{table}

\subsection{Production Deployment} 
We evaluate our framework through production deployment in our e-commerence image2product search system.
For fair comparison, the baseline method consists of the same image and text encoder with our method. The differences are 1) the baseline method obtains final doc embedding by simply averaging the image and text embedding, 2) the baseline method is trained by triplet-loss, where the label is collected from users' click behaviour.

For offline evaluation, we focus on the relevance between query images and result products.  We construct an offline evaluation set by sampling from real system log, which contains 2500 real query images and 240 thousands of candidate products, the relevance level between queries and products is labeled by human as identical/relevant/irrelevant. During offline evaluation, we use our model to extract the representation embedding of the query and products. We take the top 5 products with the highest cosine similarity to queries and  measure the identical and relevance ratio on them. 

In the online A/B test, we employed the proposed model/baseline model to extract representation vectors, which was combined with other features (item ID, item profile, user profile, user ID) and fed into the downstream CTR (Click-Through Rate) and CVR (Conversion Rate) prediction models. The CTR and CVR prediction models are four-layer MLP (Multi-Layer Perceptron) models. The CTR and CVR model were trained with simple binary cross-entropy loss on a dataset collected from online user behaviour, where click/non-click (pay/non-pay) behaviour are seen as positive/negative labels. Ultimately, we rank the candidate products by prediction CTCVR~(CTR$\times$CVR); this means that the higher the CTCVR, the higher the product is ranked in the search results.  We conducted the online A/B test during the period from 2021/12/17 to 2021/12/24.
Shown in Table~\ref{table:overall_industrial}, our method achieve a significant gain compared with the baseline model on both offline and online metric. Figure~\ref{figure:ranking_exmples} shows two examples of online retrieval, our method outperforms the base model for both easy and hard cases.

\vspace{0.2cm}
\section{Conclusion}
This paper proposes one novel framework for Image-to-Multi-Modal retrieval under industrial settings. Our method consists of the OCLEAR paradigm to address data skew and noise, and a novel model architecture for modality fusion in the doc side. Our method achieves SOTA performance on public datasets and lead to significant improvements in click-through rate and deal numbers in a real-world industrial search system. For future work, more exploration can be conducted to efficiently combine denoising and de-skew methods with our large-scale classficication paradigm.

\bibliographystyle{ACM-Reference-Format}
\bibliography{acm23}

\end{document}